\documentclass[12pt]{revtex4-1}
\usepackage{amsmath}
\usepackage{amssymb}
\usepackage[left=1cm,right=1cm]{geometry}
\usepackage[colorlinks,linkcolor=blue,citecolor=red]{hyperref}
\usepackage{feynmp}
\usepackage{slashed}
\usepackage{color}
\usepackage{graphicx}
\usepackage{makecell}
\usepackage{multirow}
\usepackage{braket}

\begin{document}

\title{\LARGE A Hamiltonian Approach to Barrier Option Pricing Under Vasicek Model}
\bigskip
\author{Chao Guo~$^{1}$}
\email{chaog@lfnu.edu.cn}
\author{Ning Yao~$^{2}$}
\email{yaoning@lfnu.edu.cn}
\affiliation{
$^{1,2}$~School of Science, Langfang Normal University, Langfang 065000, China
}
\date{\today}

\begin{abstract}

In this paper, we study option pricing under Vasicek Model by a Hamiltonian approach. Since the interest rate changes with time, we split the time to maturity into infinite steps, and the matrix element during each step could be calculated by quantum mechanics methods. Using completeness condition, the pricing kernel and the integral expression of option price could also be derived. Numerical results of option prices as functions of underlying asset price, floating rate and regression rate are also shown. 
\end{abstract}

\maketitle

%\tableofcontents\newpage
%%%%%%%%%%%%%%%%%%%%%%%%%%%%%%%%%%%%%%%%%%%%%%%%%%%%%%%%%%%%%%%%%%%%%%%%%%%%%%%%%%%%%%
\section{Introduction}
%%%%%%%%%%%%%%%%%%%%%%%%%%%%%%%%%%%%%%%%%%%%%%%%%%%%%%%%%%%%%%%%%%%%%%%%%%%%%%%%%%%%%%

In 1973, Black and Scholes~\cite{Black} applied the hedging principle developed by Merton~\cite{Merton:1973} to derive a closed-form solution for European options. Since then the field of derivative pricing has grown greatly~\cite{Amin,Rubinstein}. In resent years, path-dependent options have become increasingly popular in financial markets, and barrier options are considered to be the simplest types of path dependent options. Owing to the constraint of the barrier, barrier options give investors more flexibilities to transactions. Snyder was the first to discuss  down-and-out barrier options~\cite{Snyder}. Merton derived a closed-form solution for the down-and-out barrier call option~\cite{Merton:1973}. Chiara et al priced barrier options by a Mellin transform method~\cite{Chiara}. Karatzas and Wang gave closed-form expressions for the prices of American up-and-out put options~\cite{Karatzas}.  Baaquie et al discussed barrier options and double barrier options by path integral approach, and derived the corresponding analytical expressions~\cite{Baaquie}. Kunitomo and Ikeda studied the barrier options with floating barriers~\cite{Kunitomo}. Chen et al gave the integral expressions for floating barrier option prices by a Hamiltonian approach~\cite{Chen}. All the papers mentioned above assumed that the model parameters such as interest rate and volatility are constants. However, as we know, interest rate and volatility change with time in the real market, and the Black-Scholes model should be revised to be close to the reality. Lo et al give a closed-form expression for the price of barrier options with time dependent parameters~\cite{Lo}. Lemmens et al discussed Stochastic interest rate and volatility by means of path integral approach~\cite{Lemmens}. More about the discussions on options with time dependent parameters could be studied in~\cite{Hull,Wiggins,Roberts,Briys,Bernard}.

In this paper, we assume that the interest rate $r$ obeys the Vasicek model~\cite{Vasicek}. Using $\Delta$-arbitrage strategy, the partial differential equation (PDE) of option price $V(S,r,0)$ could be derived, which could be rewirtten into a Schr{\"o}dinger-type equation. The Hamiltonian of the option system determines the expression of the pricing kernel which could be derived by a quantum mechanical approach~\cite{Baaquie,Chen}. The pricing kernel carries all the information on option price evolution, which is in correspondence to the propagator in quantum theory. Options with floating barriers give investors more flexibilities. With the barriers moving, the opportunities for an option to touch the barriers would change, and the option price increases or decreases correspondingly. We would take up-and-out call options for instance to illustrate the effect of floating barriers and give the closed-form formula of the option price.  

Our work is organized as follows. In Section 2, we will derive the integral expressions for the barrier option price under the Vasicek model by a Hamiltonian approach. In Section 3, we will derive the integral expressions for the double barrier option price. In Section 4, We will generalize the results in Section 2 and Section 3 into the case of an option with floating barriers.
Numerical results will be discussed in Section 5. We summarize our main results in Section 6.

%%%%%%%%%%%%%%%%%%%%%%%%%%%%%%%%%%%%%%%%%%%%%%%%%%%%%%%%%%%%%%%%%%%%%%%%%%%%%%%%%%%%%%
\section{Hamiltonian Approach to Barrier Option Pricing}\label{section3}
%%%%%%%%%%%%%%%%%%%%%%%%%%%%%%%%%%%%%%%%%%%%%%%%%%%%%%%%%%%%%%%%%%%%%%%%%%%%%%%%%%%%%
Assuming that the underlying asset price $S_t$ obeys the Brownian movement
\begin{equation}
\frac{{\rm d}S_t}{S_t}=r_t{\rm d}t+\sigma_1{\rm d}W_t^1
\end{equation}
where $\sigma_1$ is the volatility of $S_t$, and $r_t$ is the stochastic interest rate which obeys the Vasicek model~\cite{Vasicek}
\begin{equation}
{\rm d}r_t=a(\theta-r_t){\rm d}t+\sigma_2{\rm d}W_t^2
\end{equation}
where $\sigma_2$ is the volatility of $r_t$, $a$ and $\theta$ indicate the regression rate and long-term mean, respectively. $W_t^1$ and $W_t^2$ are standard Brownian movements, with the covariance
\begin{equation}
{\rm Cov}({\rm d}W_t^1, {\rm d}W_t^2)=\rho {\rm d}t,\ \ |\rho|\leq 1
\end{equation}

Using the $\Delta$-arbitrage strategy, the option price $V(S, r, t)$ satisfies the follow partial differential equation (PDE)
\begin{equation}\label{optionpricePDE}
\frac{\partial V}{\partial t}+\frac{1}{2}\sigma_1^2 S^2\frac{\partial^2 V}{\partial S^2}+\sigma_1\sigma_2\rho S\frac{\partial^2 V}{\partial S\partial r}+\frac{1}{2}\sigma_2^2\frac{\partial^2 V}{\partial r^2}+rS\frac{\partial V}{\partial S}+a(\theta-r)\frac{\partial V}{\partial r}-rV=0
\end{equation}
with the barrier 
\begin{equation}\label{barrier}
U_{B}=\left\{
\begin{aligned}
0 & , & S< S_{B},\\
\infty & , & x\geq S_{B}.
\end{aligned}
\right.
\end{equation}
where $S$ is the underlying asset price, and $S_{B}$ is the barrier level. For a call option, the final condition at $t=\tau$ is
\begin{equation}
V(S,r,\tau)=(S-K)^+
\end{equation}
where $\tau$ is the time to maturity, and $K$ is the exercise price. By means of variable substitutions
\begin{equation}\label{VS}
y=\frac{S}{P(r,t;\tau)},\ \ \ \hat{V}(y,t)=\frac{V(S,r,t)}{P(r,t;\tau)}
\end{equation}
the PDE (\ref{optionpricePDE}) could be simplified into
\begin{equation}\label{simplifiedPDE}
\frac{\partial \hat{V}}{\partial t}+\frac{1}{2}\sigma^2(t)y^2\frac{\partial^2 \hat{V}}{\partial y^2}=0
\end{equation}
the final condition
\begin{equation}
    \hat{V}(y,\tau)=\frac{V(S,r,\tau)}{P(r,\tau;\tau)}=(y-K)^+
\end{equation}
where
\begin{equation}
\sigma(t)=\sqrt{\sigma_1^2+2\rho\sigma_1\sigma_2A_1(t)+\sigma_2^2A_1^2(t)}
\end{equation}
and $P(r,t;\tau)$ is the zero-coupon bond price. Under the Vasicek model, $P(r,t;\tau)$ obeys the following PDE and the final condition
\begin{equation}\begin{split}\label{VasicekPDE}
&\frac{\partial P}{\partial t}+\frac{\sigma^2}{2}\frac{\partial^2 P}{\partial r^2}+a(\theta-r)\frac{\partial P}{\partial r}-rP=0\\
&P(r,T)=1
\end{split}\end{equation}
which has the explicit solution
\begin{equation}\begin{split}
P(r,t;\tau)&=A_2(t)e^{-rA_1(t)}\\
A_1(t)&=\frac{1}{a}(1-e^{-a(\tau-t)})\\
A_2(t)&={\rm exp}\bigg[\frac{1}{a^2}[A_1^2(t)-(\tau-t)]\bigg(a^2\theta-\frac{\sigma_2^2}{2}\bigg)-\frac{\sigma_2^2}{4a}A_1^2(t)\bigg]
\end{split}\end{equation}

Making the following substitution
\begin{equation}\label{subsititution}
y=e^x,\ \ \ -\infty<x<+\infty
\end{equation}
the barrier (\ref{barrier}) could also be written as
\begin{equation}
U(x,t)=\left\{
\begin{aligned}
0 & , & x< x_B,\\
\infty & , & x\geq x_B.
\end{aligned}
\right.
\end{equation}
where
\begin{equation}
x_B={\rm ln}S_{B}-{\rm ln}A_2(t)+r(t)A_1(t)
\end{equation}
and (\ref{simplifiedPDE}) could be changed into  a Schr{\"o}dinger-type equation 
\begin{equation}
\frac{\partial \hat{V}}{\partial t}=H\hat{V}
\end{equation}
with the Hamiltonian $H$ given by
\begin{equation}
H=-\frac{1}{2}\sigma^2(t)\frac{\partial^2}{\partial x^2}+\frac{1}{2}\sigma^2(t)\frac{\partial}{\partial x}=e^{\frac{1}{2} x} H_{\rm eff} e^{-\frac{1}{2} x}+U(x)
\end{equation}
where
\begin{equation}
H_{\rm eff}=-\frac{1}{2}\sigma^2(t)\frac{\partial^2}{\partial x^2}+\frac{1}{8}\sigma^2(t)  
\end{equation}
is the effective Hamiltonian which obeys the stationary state Schr{\"o}dinger equation
\begin{equation}\label{simplifiedschrodinger}
-\frac{1}{2}\sigma^2(t)\frac{\partial^2 \phi}{\partial x^2}+\frac{1}{8}\sigma^2(t)\phi=E\phi,\ \ \ x<x_B
\end{equation}
where $\phi$ is the scattering state wave function of the option, which describes the option price at time $t$, and $E$ is corresponding to the\ ``option energy''. The solution of  (\ref{simplifiedschrodinger}) is
\begin{equation}
\phi(x,t)=\braket{x|p(t)}=e^{ip(t)(x-x_B)}-e^{-ip(t)(x-x_B)}, \ \ \ x<x_B
\end{equation}
which is the option wave function represented in coordinate space with the barrier level $x_B$, and 
\begin{equation}
p^2(t)=\frac{2E}{\sigma^2(t)}-\frac{1}{4}
\end{equation}
for $x\geq x_B$, the wave function $\phi(x,t)=0$. 

Owing to $\sigma(t)$ changes with time, we discretize $\tau$ so that there are $N$ steps to maturity, with each time step $\epsilon=\tau/N$. $\sigma(t)$ tends to be a constant during each small enough time step. The pricing kernel could be denoted as
\begin{equation}\begin{split}
p_{\rm BO}(x,x^\prime;\tau)&=\braket{x|e^{-\tau H}|x^\prime}\\
&=\lim_{\epsilon\to 0}\int_{-\infty}^{x_{B,1}} {\rm d}x_1\int_{-\infty}^{x_{B,2}} {\rm d}x_2...\int_{-\infty}^{x_{B,N-1}} {\rm d}x_{N-1}\braket{x|e^{-\tau H}|x_1}\braket{x_1|e^{-\tau H}|x_2}...\braket{x_{N-1}|e^{-\tau H}|x^\prime}
\end{split}\end{equation}
where the completeness condition 
\begin{equation}
\int_{-\infty}^{x_{B,j}} {\rm d}x_j \ket{x_j}\bra{x_j}=1
\end{equation}
has been used. The $(j+1)th$ matrix element is
\begin{equation}\begin{split}\label{jthmatrix}
\braket{x_j|e^{-\epsilon H}|x_{j+1}}=&\int_0^{+\infty}\frac{{\rm d}p_j}{2\pi}\braket{x_j|e^{-\epsilon H}|p_j}\braket{p_j|x_{j+1}}\\
=&e^{\alpha(x_j-x_{j+1})}e^{-\epsilon\gamma_j}\int_{0}^{+\infty}\frac{{\rm{d}} p_j}{2\pi}e^{-\frac{1}{2}\epsilon\sigma_j^2 p_j^2}\big[e^{ip_j(x_j-x_{B,j})}-e^{-ip_j(x_j-x_{B,j})}\big]\\
\times&\big[e^{-ip_j(x_{j+1}-x_{B,j+1})}-e^{ip_j(x_{j+1}-x_{B,j+1})}\big]\\
=&e^{\alpha(x_j-x_{j+1})}e^{-\epsilon\gamma_j}\int_{-\infty}^{+\infty}\frac{{\rm{d}} p_j}{2\pi}e^{-\frac{1}{2}\epsilon\sigma_j^2 p_j^2}\big[e^{ip_j(x_j-x_{j+1}-x_{B,j}+x_{B,j+1})}-e^{ip_j(x_j+x_{j+1}-x_{B,j}-x_{B,j+1})}\big]
\end{split}\end{equation}

Using Gaussian integral formula, the integral in (\ref{jthmatrix}) could be calculated as
\begin{equation}\begin{split}
&\int_{-\infty}^{+\infty}\frac{{\rm{d}} p_j}{2\pi}e^{-\frac{1}{2}\epsilon\sigma_j^2 p_j^2}\big[e^{ip_j(x_j-x_{j+1}-x_{B,j}+x_{B,j+1})}-e^{ip_j(x_j+x_{j+1}-x_{B,j}-x_{B,j+1})}\big]\\
=&\int_{-\infty}^{+\infty}\frac{{\rm{d}} p_j}{2\pi}e^{-\frac{1}{2}\epsilon\sigma_j^2\big[p_j-\frac{i(x_j-x_{j+1}-x_{B,j}+x_{B,j+1})}{\epsilon\sigma_j^2}\big]^2}e^{-\frac{(x_j-x_{j+1}-x_{B,j}+x_{B,j+1})^2}{2\epsilon\sigma_j^2}}\\
-&\int_{-\infty}^{+\infty}\frac{{\rm{d}} p_j}{2\pi}e^{-\frac{1}{2}\epsilon\sigma_j^2\big[p_j-\frac{i(x_j+x_{j+1}-x_{B,j}-x_{B,j+1})}{\epsilon\sigma_j^2}\big]^2}e^{-\frac{(x_j+x_{j+1}-x_{B,j}-x_{B,j+1})^2}{2\epsilon\sigma_j^2}}\\
=&\frac{1}{\sqrt{2\pi\epsilon\sigma_j^2}}\bigg[e^{-\frac{(x_j-x_{j+1}-x_{B,j}+x_{B,j+1})^2}{2\epsilon\sigma_j^2}}-e^{-\frac{(x_j+x_{j+1}-x_{B,j}-x_{B,j+1})^2}{2\epsilon\sigma_j^2}}\bigg]
\end{split}\end{equation}
Similarly,
\begin{equation}\begin{split}\label{2matrix}
\braket{x_{j-1}|e^{-2\epsilon H}|x_{j+1}}=&\int_{-\infty}^{x_{B,j}} {\rm d}x_j \braket{x_{j-1}|e^{-\epsilon H}|x_{j}}\braket{x_j|e^{-\epsilon H}|x_{j+1}}\\
=&\int_{-\infty}^{x_{B,j}} {\rm d}x_j e^{\frac{1}{2}(x_{j-1}-x_{j+1})}e^{-\epsilon(\gamma_{j-1}+\gamma_j)}\frac{1}{\sqrt{2\pi\epsilon\sigma_{j-1}^2}}\frac{1}{\sqrt{2\pi\epsilon\sigma_{j}^2}}\\
\times&\bigg[e^{-\frac{1}{2\epsilon\sigma_{j-1}^2}(x_j-\mathcal{A})^2}-e^{-\frac{1}{2\epsilon\sigma_{j-1}^2}(x_j+\mathcal{C})^2}\bigg]\bigg[e^{-\frac{1}{2\epsilon\sigma_{j}^2}(x_{j}-\mathcal{B})^2}-e^{-\frac{1}{2\epsilon\sigma_{j}^2}(x_{j}+\mathcal{D})^2}\bigg]
\end{split}\end{equation}
where
\begin{equation}\begin{split}
\mathcal{A}&=x_{j-1}-x_{B,j-1}+x_{B,j},\ \ \mathcal{B}=x_{j+1}+x_{B,j}-x_{B,j+1}\\
\mathcal{C}&=x_{j-1}-x_{B,j-1}-x_{B,j},\ \ \mathcal{D}=x_{j+1}-x_{B,j}-x_{B,j+1}
\end{split}\end{equation}
now we calculate the four integrals in (\ref{2matrix}) 
\begin{equation}\begin{split}
&\int_{-\infty}^{x_{B,j}} {\rm d}x_j e^{-\frac{1}{2\epsilon\sigma_{j-1}^2}(x_j-\mathcal{A})^2-\frac{1}{2\epsilon\sigma_j^2}(x_j-\mathcal{B})^2}\\
=&\int_{-\infty}^{x_{B,j}} {\rm d}x_j e^{-\frac{\sigma_{j-1}^2+\sigma_j^2}{2\epsilon\sigma_{j-1}^2\sigma_j^2}\big[x_j-\frac{\mathcal{A}\sigma_j^2+\mathcal{B}\sigma_{j-1}^2}{\sigma_{j-1}^2+\sigma_j^2}\big]^2}e^{-\frac{(\mathcal{A}-\mathcal{B})^2}{2\epsilon(\sigma_{j-1}^2+\sigma_j^2)}}\\
=&\frac{1}{2}\sqrt{\frac{2\pi\epsilon\sigma_{j-1}^2\sigma_j^2}{\sigma_{j-1}^2+\sigma_j^2}}\bigg(1+ {\rm erf}\bigg[\frac{x_{B,j}(\sigma_{j-1}^2+\sigma_j^2)-\mathcal{A}\sigma_j^2-\mathcal{B}\sigma_{j-1}^2}{\sqrt{2\epsilon\sigma_{j-1}^2\sigma_j^2(\sigma_{j-1}^2+\sigma_j^2)}}\bigg]\bigg)e^{-\frac{(\mathcal{A}-\mathcal{B})^2}{2\epsilon(\sigma_{j-1}^2+\sigma_j^2)}}\\
=&\frac{1}{2}\sqrt{\frac{2\pi\epsilon\sigma_{j-1}^2\sigma_j^2}{\sigma_{j-1}^2+\sigma_j^2}}\bigg(1+ {\rm erf}\bigg[\frac{-(x_{j-1}-x_{B,j-1})\sigma_j^2-(x_{j+1}-x_{B,j+1})\sigma_{j-1}^2}{\sqrt{2\epsilon\sigma_{j-1}^2\sigma_j^2(\sigma_{j-1}^2+\sigma_j^2)}}\bigg]\bigg)e^{-\frac{(\mathcal{A}-\mathcal{B})^2}{2\epsilon(\sigma_{j-1}^2+\sigma_j^2)}}
\end{split}\end{equation}
\begin{equation}\begin{split}
&\int_{-\infty}^{x_{B,j}} {\rm d}x_j e^{-\frac{1}{2\epsilon\sigma_{j-1}^2}(x_j+\mathcal{C})^2-\frac{1}{2\epsilon\sigma_j^2}(x_j+\mathcal{D})^2}\\
=&\int_{-\infty}^{x_{B,j}} {\rm d}x_j e^{-\frac{\sigma_{j-1}^2+\sigma_j^2}{2\epsilon\sigma_{j-1}^2\sigma_j^2}\big[x_j+\frac{\mathcal{C}\sigma_j^2+\mathcal{D}\sigma_{j-1}^2}{\sigma_{j-1}^2+\sigma_j^2}\big]^2}e^{-\frac{(\mathcal{C}-\mathcal{D})^2}{2\epsilon(\sigma_{j-1}^2+\sigma_j^2)}}\\
=&\frac{1}{2}\sqrt{\frac{2\pi\epsilon\sigma_{j-1}^2\sigma_j^2}{\sigma_{j-1}^2+\sigma_j^2}}\bigg(1+ {\rm erf}\bigg[\frac{x_{B,j}(\sigma_{j-1}^2+\sigma_j^2)+\mathcal{C}\sigma_j^2+\mathcal{D}\sigma_{j-1}^2}{\sqrt{2\epsilon\sigma_{j-1}^2\sigma_j^2(\sigma_{j-1}^2+\sigma_j^2)}}\bigg]\bigg)e^{-\frac{(\mathcal{C}-\mathcal{D})^2}{2\epsilon(\sigma_{j-1}^2+\sigma_j^2)}}\\
=&\frac{1}{2}\sqrt{\frac{2\pi\epsilon\sigma_{j-1}^2\sigma_j^2}{\sigma_{j-1}^2+\sigma_j^2}}\bigg(1+ {\rm erf}\bigg[\frac{(x_{j-1}-x_{B,j-1})\sigma_j^2+(x_{j+1}-x_{B,j+1})\sigma_{j-1}^2}{\sqrt{2\epsilon\sigma_{j-1}^2\sigma_j^2(\sigma_{j-1}^2+\sigma_j^2)}}\bigg]\bigg)e^{-\frac{(\mathcal{C}-\mathcal{D})^2}{2\epsilon(\sigma_{j-1}^2+\sigma_j^2)}}
\end{split}\end{equation}
\begin{equation}\begin{split}
&\int_{-\infty}^{x_{B,j}} {\rm d}x_j e^{-\frac{1}{2\epsilon\sigma_{j-1}^2}(x_j-\mathcal{A})^2-\frac{1}{2\epsilon\sigma_j^2}(x_j+\mathcal{D})^2}\\
=&\int_{-\infty}^{x_{B,j}} {\rm d}x_j e^{-\frac{\sigma_{j-1}^2+\sigma_j^2}{2\epsilon\sigma_{j-1}^2\sigma_j^2}\big[x_j-\frac{\mathcal{A}\sigma_j^2-\mathcal{D}\sigma_{j-1}^2}{\sigma_{j-1}^2+\sigma_j^2}\big]^2}e^{-\frac{(\mathcal{A}+\mathcal{D})^2}{2\epsilon(\sigma_{j-1}^2+\sigma_j^2)}}\\
=&\frac{1}{2}\sqrt{\frac{2\pi\epsilon\sigma_{j-1}^2\sigma_j^2}{\sigma_{j-1}^2+\sigma_j^2}}\bigg(1+ {\rm erf}\bigg[\frac{x_{B,j}(\sigma_{j-1}^2+\sigma_j^2)-\mathcal{A}\sigma_j^2+\mathcal{D}\sigma_{j-1}^2}{\sqrt{2\epsilon\sigma_{j-1}^2\sigma_j^2(\sigma_{j-1}^2+\sigma_j^2)}}\bigg]\bigg)e^{-\frac{(\mathcal{A}+\mathcal{D})^2}{2\epsilon(\sigma_{j-1}^2+\sigma_j^2)}}\\
=&\frac{1}{2}\sqrt{\frac{2\pi\epsilon\sigma_{j-1}^2\sigma_j^2}{\sigma_{j-1}^2+\sigma_j^2}}\bigg(1+ {\rm erf}\bigg[\frac{-(x_{j-1}-x_{B,j-1})\sigma_j^2+(x_{j+1}-x_{B,j+1})\sigma_{j-1}^2}{\sqrt{2\epsilon\sigma_{j-1}^2\sigma_j^2(\sigma_{j-1}^2+\sigma_j^2)}}\bigg]\bigg)e^{-\frac{(\mathcal{A}+\mathcal{D})^2}{2\epsilon(\sigma_{j-1}^2+\sigma_j^2)}}
\end{split}\end{equation}
\begin{equation}\begin{split}
&\int_{-\infty}^{x_{B,j}} {\rm d}x_j e^{-\frac{1}{2\epsilon\sigma_{j-1}^2}(x_j+\mathcal{C})^2-\frac{1}{2\epsilon\sigma_j^2}(x_j-\mathcal{B})^2}\\
=&\int_{-\infty}^{x_{B,j}} {\rm d}x_j e^{-\frac{\sigma_{j-1}^2+\sigma_j^2}{2\epsilon\sigma_{j-1}^2\sigma_j^2}\big[x_j-\frac{\mathcal{C}\sigma_j^2-\mathcal{B}\sigma_{j-1}^2}{\sigma_{j-1}^2+\sigma_j^2}\big]^2}e^{-\frac{(\mathcal{B}+\mathcal{C})^2}{2\epsilon(\sigma_{j-1}^2+\sigma_j^2)}}\\
=&\frac{1}{2}\sqrt{\frac{2\pi\epsilon\sigma_{j-1}^2\sigma_j^2}{\sigma_{j-1}^2+\sigma_j^2}}\bigg(1+ {\rm erf}\bigg[\frac{x_{B,j}(\sigma_{j-1}^2+\sigma_j^2)+\mathcal{C}\sigma_j^2-\mathcal{B}\sigma_{j-1}^2}{\sqrt{2\epsilon\sigma_{j-1}^2\sigma_j^2(\sigma_{j-1}^2+\sigma_j^2)}}\bigg]\bigg)e^{-\frac{(\mathcal{B}+\mathcal{C})^2}{2\epsilon(\sigma_{j-1}^2+\sigma_j^2)}}\\
=&\frac{1}{2}\sqrt{\frac{2\pi\epsilon\sigma_{j-1}^2\sigma_j^2}{\sigma_{j-1}^2+\sigma_j^2}}\bigg(1+ {\rm erf}\bigg[\frac{(x_{j-1}-x_{B,j-1})\sigma_j^2-(x_{j+1}-x_{B,j+1})\sigma_{j-1}^2}{\sqrt{2\epsilon\sigma_{j-1}^2\sigma_j^2(\sigma_{j-1}^2+\sigma_j^2)}}\bigg]\bigg)e^{-\frac{(\mathcal{B}+\mathcal{C})^2}{2\epsilon(\sigma_{j-1}^2+\sigma_j^2)}}
\end{split}\end{equation}
where
\begin{equation}
{\rm erf}(x)=\frac{2}{\sqrt{\pi}}\int_0^x e^{-\eta^2}{\rm d}\eta
\end{equation}
is the error function. Owning to the error function is an odd function, (\ref{2matrix}) could be simplified into
\begin{equation}\begin{split}
\braket{x_{j-1}|e^{-2\epsilon H}|x_{j+1}}=&\frac{e^{\frac{1}{2}(x_{j-1}-x_{j+1})}e^{-\frac{1}{8}\epsilon(\sigma_{j-1}^2+\sigma_j^2)}}{\sqrt{2\pi\epsilon(\sigma_{j-1}^2+\sigma_j^2)}}\bigg[e^{-\frac{(\mathcal{A}-\mathcal{B})^2}{2\epsilon(\sigma_{j-1}^2+\sigma_j^2)}}-e^{-\frac{(\mathcal{B}+\mathcal{C})^2}{2\epsilon(\sigma_{j-1}^2+\sigma_j^2)}}\bigg]\\
=&\frac{e^{\frac{1}{2}(x_{j-1}-x_{j+1})}e^{-\frac{1}{8}\epsilon(\sigma_{j-1}^2+\sigma_j^2)}}{\sqrt{2\pi\epsilon(\sigma_{j-1}^2+\sigma_j^2)}}\bigg[e^{-\frac{(x_{j-1}-x_{j+1}-x_{B,j-1}+x_{B,j+1})^2}{2\epsilon(\sigma_{j-1}^2+\sigma_j^2)}}-e^{-\frac{(x_{j-1}+x_{j+1}-(x_{B,j-1}+x_{B,j+1}))^2}{2\epsilon(\sigma_{j-1}^2+\sigma_j^2)}}\bigg]
\end{split}\end{equation}

Repeating the above calculation, the pricing kernel is
\begin{equation}
\braket{x|e^{-\tau H}|x^{\prime}}=\frac{e^{\frac{1}{2}(x-x^\prime)}e^{-\frac{1}{8}\lim\limits_{\epsilon\to 0}\epsilon\sum_{j=1}^N \sigma_j^2}}{\sqrt{2\pi\lim\limits_{\epsilon\to 0}\epsilon\sum_{j=1}^N \sigma_j^2}}\bigg[e^{-\frac{[(x-x^\prime)-(x_{B,0}-x_{B,\tau})]^2}{2\lim\limits_{\epsilon\to 0}\epsilon\sum_{j=1}^N \sigma_j^2}}-e^{-\frac{[(x+x^\prime)-(x_{B,0}+x_{B,\tau})]^2}{2\lim\limits_{\epsilon\to 0}\epsilon\sum_{j=1}^N \sigma_j^2}}\bigg]
\end{equation}
where
\begin{equation}\begin{split}
\lim_{\epsilon\to 0}\epsilon\sum_{j=1}^N\sigma_j^2\equiv\sigma_{\rm eff}&=\epsilon\sum_{j=1}^N[\sigma_1^2+2\rho\sigma_1\sigma_2 A_1(t)+\sigma_2^2 A_1^2(t)]^2\\
&=(\sigma_1^2+\frac{2\rho}{a}\sigma_1\sigma_2+\frac{\sigma_2^2}{a^2})\tau-\frac{2\sigma_2}{a^2}(\rho\sigma_1+\frac{\sigma_2}{a})(1-e^{-a\tau})+\frac{\sigma_2^2}{2a^3}(1-e^{-2a\tau})
\end{split}\end{equation}
could be viewed as the effective volatility, and 
\begin{equation}\begin{split}
   x_{B,0}=&{\rm ln}S_B-{\rm ln}A_2(0)+rA_1(0)\\
   x_{B,\tau}=&{\rm ln}S_B-{\rm ln}A_2(\tau)+rA_1(\tau)
\end{split}\end{equation}
the option price could be denoted as 
\begin{equation}
V(S,r;0)=P(r,0;\tau)\int_{{\rm ln}K}^{x_{B,\tau}}\braket{x|e^{-\tau H}|x^\prime}(e^{x^\prime}-K){\rm d}x^\prime
\end{equation}

%%%%%%%%%%%%%%%%%%%%%%%%%%%%%%%%%%%%%%%%%%%%%%%%%%%%%%%%%%%%%%%%%%%%%%%%%%%%%%%%%%%%%%
\section{Hamiltonian Approach to Double Barrier Option Pricing}
%%%%%%%%%%%%%%%%%%%%%%%%%%%%%%%%%%%%%%%%%%%%%%%%%%%%%%%%%%%%%%%%%%%%%%%%%%%%%%%%%%%%%

The double barrier option Hamiltonian is
\begin{equation}\begin{split}
H&=-\frac{1}{2}\sigma^2(t)\frac{\partial^2}{\partial x^2}+\frac{1}{2}\sigma^2(t)\frac{\partial}{\partial x}+U(x)\\
&=e^{\frac{1}{2}x}\bigg(-\frac{1}{2}\sigma^2(t)\frac{\partial^2}{\partial x^2}\bigg)e^{-\frac{1}{2}x}+\frac{1}{8}\sigma^2(t)+U(x)\\
&\equiv e^{\frac{1}{2}x}H_{\rm eff}e^{-\frac{1}{2}x}+\frac{1}{8}\sigma^2(t)+U(x)
\end{split}\end{equation}
with the potential $U(x)$
\begin{equation}
U(x)=\left\{
\begin{aligned}
\infty & , & x\leq x_{B_1},\\
0 & , & x_{B_1}<x<x_{B_2},\\
\infty & , & x\geq x_{B_2}.
\end{aligned}
\right.
\end{equation}
where
\begin{equation}\begin{split}
    x_{B_1}=&{\rm ln}S_{B_1}-{\rm ln}A_2(t)+rA_1(t)\\
    x_{B_2}=&{\rm ln}S_{B_2}-{\rm ln}A_2(t)+rA_1(t)
\end{split}\end{equation}
are the lower and higher barrier levels, respectively. The Schr{\"o}dinger equation for $H_{\rm eff}$ is 
\begin{equation}
-\frac{1}{2}\sigma^2(t)\frac{\partial^2 \phi}{\partial x^2}=E\phi,\ \ \ x_{B_1}<x<x_{B_2}
\end{equation}
with the eigenstate
\begin{equation}
 \phi_n(x)=\left\{
\begin{aligned}
\sqrt{\frac{2}{x_{B_2}-x_{B_1}}}\sin&[p_n(x-x_{B_1})]  , & x_{B_1}<x<x_{B_2},\\
0 & , & x\leq x_{B_1},\ x\geq x_{B_2}.
\end{aligned}
\right.   
\end{equation}
where
\begin{equation}
p_n=\frac{n\pi}{x_{B_2}-x_{B_1}}=\frac{n\pi}{{\rm ln}S_{B_2}-{\rm ln}S_{B_1}},\ \ E_n=\frac{1}{2}\sigma^2(t)p_n^2, \ \ n=1,2,3,...
\end{equation}

Owing to $\sigma$ changes with time, we also discretize $\tau$ to $N$ steps, with each step $\epsilon=\tau/N$. The $(j+1)th$ matrix element is
\begin{equation}\begin{split}
\braket{x_j|e^{-\epsilon H}|x_{j+1}}&=\braket{x_j|e^{e^{\frac{1}{2}x}\big(\frac{1}{2}\epsilon \sigma^2(t)\frac{\partial^2}{\partial x^2}\big)e^{-\frac{1}{2}x}-\frac{1}{8}\epsilon\sigma^2(t)}|x_{j+1}}\\
&=e^{\frac{1}{2}(x_j-x_{j+1})}e^{-\frac{1}{8}\epsilon\sigma_j^2}\sum_{n=1}^\infty \braket{x_j|e^{-\frac{1}{2}\epsilon\sigma^2(t)\hat{p}^2}|n}\braket{n|x_{j+1}}\\
&=e^{\frac{1}{2}(x_j-x_{j+1})}e^{-\frac{1}{8}\epsilon\sigma_j^2}\sum_{n=1}^\infty e^{-\frac{1}{2}\epsilon p_{n}^2\sigma_j^2}\phi_n(x_j)\phi_n(x_{j+1})
\end{split}\end{equation}
where $\hat{p}=-i\frac{\partial}{\partial x}$ is the momentum operator, and
\begin{equation}
\braket{x_j|n}=\phi_n(x_j)=\sqrt{\frac{2}{x_{B_2}-x_{B_1}}}\sin{p_{n}(x_j-x_{B_{1,j}})}
\end{equation}
similarly,
\begin{equation}\begin{split}
\braket{x_j|e^{-2\epsilon H}|x_{j+2}}&=\int_{x_{B_1}}^{x_{B_2}}{\rm d}x_{j+1}\braket{x_j|e^{-\epsilon H}|x_{j+1}}\braket{x_{j+1}|e^{-\epsilon H}|x_{j+2}}\\
&=\sum_{n=1}^{\infty}\sum_{n^\prime=1}^{\infty}e^{\frac{1}{2}(x_j-x_{j+2})}e^{-\frac{1}{8}\epsilon(\sigma_j^2+\sigma_{j+1}^2)}\phi_n(x_j)\bigg[\int_{x_{B_1}}^{x_{B_2}}{\rm d}x_{j+1}\phi_n(x_{j+1})\phi_{n^\prime}(x_{j+1})\bigg]\phi_{n^\prime}(x_{j+2})\\
&=e^{\frac{1}{2}(x_j-x_{j+2})}e^{-\frac{1}{8}\epsilon(\sigma_j^2+\sigma_{j+1}^2)}\sum_{n=1}^{\infty}e^{-\frac{1}{2}\epsilon p_{n}^2(\sigma_j^2+\sigma_{j+1}^2)}\phi_n(x_j)\phi_n(x_{j+2})
\end{split}\end{equation}
where the orthonormalization condition 
\begin{equation}
    \int_{x_{B_1}}^{x_{B_2}}{\rm d}x\  \phi_n(x)\phi_{n^\prime}(x)=\delta_{nn^\prime}=\left\{
\begin{aligned}
0 & , & n \neq n^\prime,\\
1 & , & n=n^\prime .
\end{aligned}
\right.
\end{equation}
has been used. Repeating the above calculation, the pricing kernel for double barrier option is
\begin{equation}\label{kernel}
\braket{x|e^{-\tau H}|x^{\prime}}=e^{\frac{1}{2}(x-x^\prime)}e^{-\frac{1}{8}\lim\limits_{\epsilon\to 0}\epsilon\sum_{j=1}^N\sigma_j^2}\sum_{n=1}^{\infty}e^{-\frac{1}{2}\epsilon p_{n}^2\lim\limits_{\epsilon\to 0}\sum_{j=1}^N\sigma_j^2}\phi_n(x)\phi_n(x^\prime)
\end{equation}
where
\begin{equation}\begin{split}
\phi_n(x)&=\sqrt{\frac{2}{{\rm ln}S_{B_2}-{\rm ln}S_{B_1}}}\sin{p_n[x-x_{B_1}(0)]}\\
\phi_n(x^\prime)&=\sqrt{\frac{2}{{\rm ln}S_{B_2}-{\rm ln}S_{B_1}}}\sin{p_n[x^\prime-x_{B_1}(\tau)]}\\
x_{B_1}(0)&={\rm ln}S_{B_1}-{\rm ln}A_2(0)+rA_1(0)\\
x_{B_1}(\tau)&={\rm ln}S_{B_1}-{\rm ln}A_2(\tau)+rA_1(\tau)
\end{split}\end{equation}
and the option price is
\begin{equation}
V(S,r;0)=P(r,0;\tau)\int_{{\rm ln}K}^{x_{B_2}(\tau)}\braket{x|e^{-\tau H}|x^\prime}(e^{x^\prime}-K){\rm d}x^\prime
\end{equation}
%%%%%%%%%%%%%%%%%%%%%%%%%%%%%%%%%%%%%%%%%%%%%%%%%%%%%%%%%%%%%%%%%%%%%%%%
\section{Option Pricing with Floating Barriers}
%%%%%%%%%%%%%%%%%%%%%%%%%%%%%%%%%%%%%%%%%%%%%%%%%%%%%%

Consider an option with the following exponential boundaries~\cite{Kunitomo}
\begin{equation}
S_Be^{\delta t},\ \ \ t\in[0,\tau]
\end{equation}
where $S_B$ is the initial barrier level, and $\delta$ is the floating rate, which denotes the change rate of the barrier. Using substitution (\ref{subsititution}), the barrier could be rewritten as
\begin{equation}
x_B={\rm ln}S_B-{\rm ln}A_2(t)+rA_1(t)+\delta t,\ \ \ t\in[0,\tau]
\end{equation}
and the $j$th matrix element is
\begin{equation}
\braket{x_j|e^{-\epsilon H}|x_{j+1}}=e^{\alpha(x_j-x_{j+1})}e^{-\epsilon\gamma_j}\int_{-\infty}^{+\infty}\frac{{\rm d}p_j}{2\pi}e^{-\frac{1}{2}\epsilon\sigma_j^2 p_j^2}\big[e^{ip_j(x_j-x_{j+1}-x_{B,j}+x_{B,{j+1}})}-e^{ip_j(x_j+x_{j+1}-x_{B,j}-x_{B,{j+1}})}\big]
\end{equation}
where
\begin{equation}
x_{B,j}={\rm ln}S_B-{\rm ln}A_2(j\epsilon)+rA_1(j\epsilon)+\delta j\epsilon
\end{equation}
is the barrier at $t=j\epsilon$. After a series of calculations, the price kernel could be denoted as
 \begin{equation}
 \braket{x|e^{-\tau H}|x^{\prime}}=\frac{e^{\frac{1}{2}(x-x^\prime)}e^{-\frac{1}{8}\lim\limits_{\epsilon\to 0}\epsilon\sum_{j=1}^N \sigma_j^2}}{\sqrt{2\pi\lim\limits_{\epsilon\to 0}\epsilon\sum_{j=1}^N \sigma_j^2}}\bigg[e^{-\frac{[(x-x^\prime)-(x_{B,0}-x_{B,\tau})]^2}{2\lim\limits_{\epsilon\to 0}\epsilon\sum_{j=1}^N \sigma_j^2}}-e^{-\frac{[(x+x^\prime)-(x_{B,0}+x_{B,\tau})]^2}{2\lim\limits_{\epsilon\to 0}\epsilon\sum_{j=1}^N \sigma_j^2}}\bigg]
\end{equation}
where
\begin{equation}\begin{split}
    x_{B,0}=&{\rm ln}S_B-{\rm ln}A_2(0)+rA_1(0)\\
    x_{B,\tau}=&{\rm ln}S_B-{\rm ln}A_2(\tau)+rA_1(\tau)+\delta\tau
\end{split}\end{equation}
and the option price is
\begin{equation}
V(x,r;0)=P(r;0,\tau)\int_{{\rm ln}K}^{x_{B,\tau}}\braket{x|e^{-\tau H}|x^\prime}(e^{x^\prime}-K){\rm d}x^\prime
\end{equation}

For a double floating barrier option, the lower bound and the upper bound could be denoted as
\begin{equation}\begin{split}
x_{B_1}(t)=&{\rm ln}S_{B_1}-{\rm ln}A_2(t)+rA_1(t)-\delta t\\
x_{B_2}(t)=&{\rm ln}S_{B_2}-{\rm ln}A_2(t)+rA_1(t)+\delta t,\ \ \ \ t\in[0,\tau]
\end{split}\end{equation}
respectively, which indicates that the two barriers either get far away from each other ($\delta>0$), or get close to each other ($\delta<0$). Similar to (\ref{kernel}), the pricing kernel for a double barrier option with floating barriers could be calculated step by step as
 \begin{equation}
 \braket{x|e^{-\tau H}|x^{\prime}}=e^{\frac{1}{2}(x-x^\prime)}e^{-\frac{1}{8}\lim\limits_{\epsilon\to 0}\epsilon\sum_{j=1}^N\sigma_j^2}\sum_{n=1}^{\infty}e^{-\frac{1}{2}\lim\limits_{\epsilon\to 0}\epsilon \sum_{j=1}^N p_{n,j}^2\sigma_j^2}\phi_n(x,0)\phi_n(x^\prime,\tau)
 \end{equation}
 where
 \begin{equation}\begin{split}
 p_{n,j}&=\frac{n\pi}{{\rm ln}S_{B_2}-{\rm ln}S_{B_1}+2j\delta\epsilon}\\
 \phi_n(x,t)&=\sqrt{\frac{2}{{\rm ln}S_{B_2}-{\rm ln}S_{B_1}+2\delta t}}\sin{[p_{n,j}(x-x_{B_1})]}
 \end{split} \end{equation}
and 
\begin{equation}\begin{split}
\sum_{j=1}^N p_{n,j}^2\sigma_j^2&=n^2\pi^2\sum_{j=1}^N\frac{\sigma_1^2+2\rho\sigma_1\sigma_2A_1(j\epsilon)+\sigma_2^2A_1(j\epsilon)^2}{({\rm ln}S_{B_2}-{\rm ln}S_{B_1}+2j\delta\epsilon)^2}\\
&=\frac{1}{4\delta^2\epsilon}\bigg[\bigg(\sigma_1^2+\frac{2\rho}{a}\sigma_1\sigma_2+\frac{\sigma_2^2}{a^2}\bigg)\sum_{j=0}^N\frac{1}{(j+\frac{{\rm ln}S_{B_2}-{\rm ln}S_{B_1}}{2\delta\epsilon})^2}\\
&-\frac{2\sigma_2}{a}\bigg(\frac{\sigma_2}{a}+\rho\sigma_1\bigg)e^{-a\tau}\sum_{j=0}^N\frac{e^{ja\epsilon}}{(j+\frac{{\rm ln}S_{B_2}-{\rm ln}S_{B_1}}{2\delta\epsilon})^2}+\frac{\sigma_2^2}{a^2}e^{-2a\tau}\sum_{j=0}^N\frac{e^{2ja\epsilon}}{(j+\frac{{\rm ln}S_{B_2}-{\rm ln}S_{B_1}}{2\delta\epsilon})^2}\bigg]\\
&=\frac{1}{4\delta^2\epsilon}\bigg[\bigg(\sigma_1^2+\frac{2\rho}{a}\sigma_1\sigma_2+\frac{\sigma_2^2}{a^2}\bigg)\bigg(\psi^\prime\bigg(\frac{{\rm ln}S_{B_2}-{\rm ln}S_{B_1}}{2\delta\epsilon}\bigg)-\psi^\prime\bigg(\frac{{\rm ln}S_{B_2}-{\rm ln}S_{B_1}}{2\delta\epsilon}+N+1\bigg)\bigg)\\
&-\frac{2\sigma_2}{a}\bigg(\frac{\sigma_2}{a}+\rho\sigma_1\bigg)e^{-a\tau}\bigg({\rm L}\bigg(e^{a\epsilon},2,\frac{{\rm ln}S_{B_2}-{\rm ln}S_{B_1}}{2\delta\epsilon}\bigg)-{\rm L}\bigg(e^{a\epsilon},2,\frac{{\rm ln}S_{B_2}-{\rm ln}S_{B_1}}{2\delta\epsilon}+N+1\bigg)\bigg)\\
&+\frac{\sigma_2^2}{a^2}e^{-2a\tau}\bigg({\rm L}\bigg(e^{2a\epsilon},2,\frac{{\rm ln}S_{B_2}-{\rm ln}S_{B_1}}{2\delta\epsilon}\bigg)-{\rm L}\bigg(e^{2a\epsilon},2,\frac{{\rm ln}S_{B_2}-{\rm ln}S_{B_1}}{2\delta\epsilon}+N+1\bigg)\bigg)\bigg]
\end{split}\end{equation}
where
\begin{equation}
\psi(z)=\frac{\Gamma^\prime(z)}{\Gamma(z)}
\end{equation}
is the $\psi$ function, and
\begin{equation}
\Gamma(z)=\int_0^{+\infty}e^{-t}t^{z-1}{\rm d}t,\ \ \ {\rm Re}z>0
\end{equation}
is the $\Gamma$ function,
\begin{equation}
{\rm L}(z,s,a)=\sum_{n=0}^{+\infty}\frac{z^n}{(a+n)^s}
\end{equation}
is the Lerch zeta function.

The option price could be denoted as
\begin{equation}
V(x,r;0)=P(r;0,\tau)\int_{{\rm ln}K}^{x_{B_2}(\tau)}\braket{x|e^{-\tau H}|x^\prime}(e^{x^\prime}-K){\rm d}x^\prime
\end{equation}
%%%%%%%%%%%%%%
\section{Numerical Results}
\begin{figure}[!htbp]
\begin{center}
\includegraphics[width=0.49\linewidth]{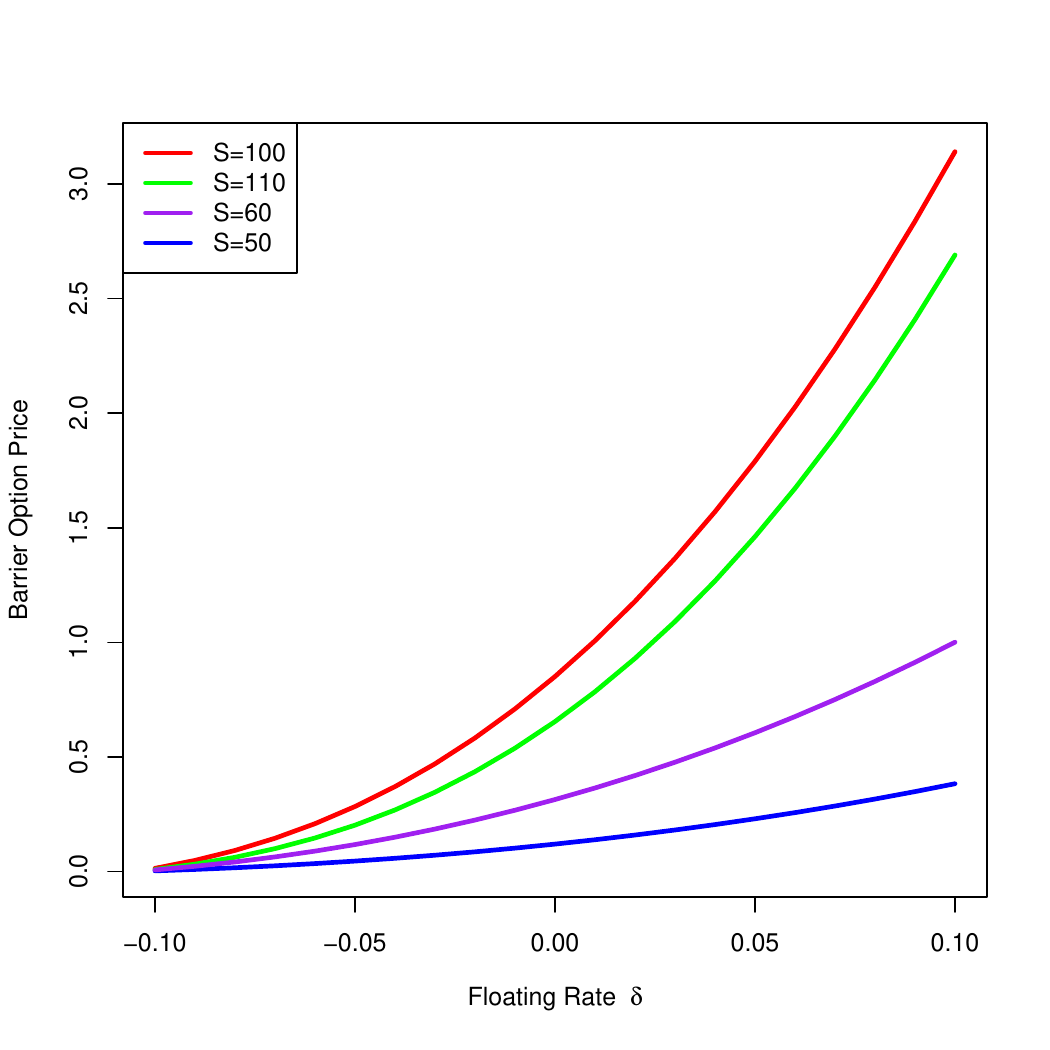}
\includegraphics[width=0.49\linewidth]{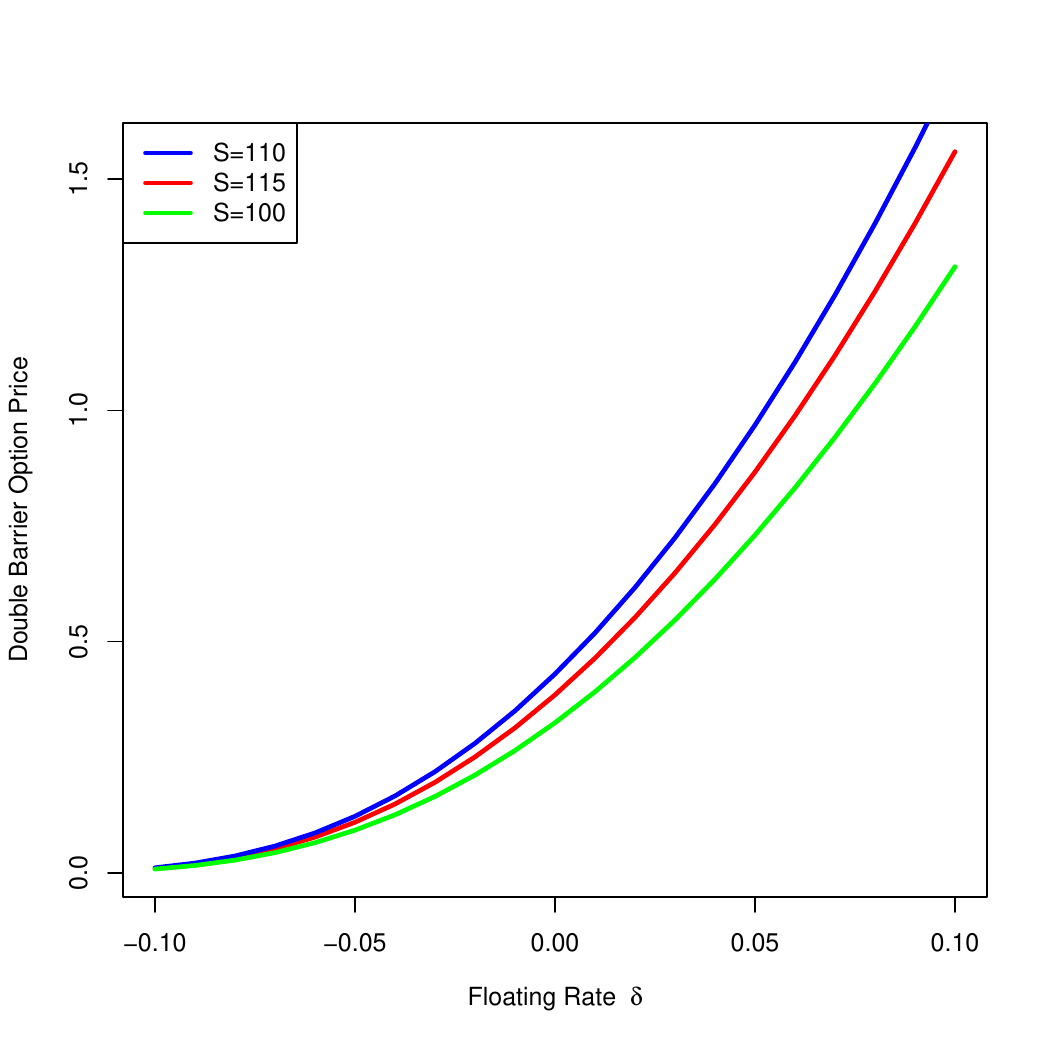}
\end{center}
\caption{Single barrier up-and-out call price (left) and  double barrier up-and-out call price (right) as functions of floating rates $\delta$. Parameters: $S_{B_1}=100$, $S_B=S_{B_2}=130$, $K=100$, $\rho=0.5$, $a=1$, $\theta=0.04$, $r_0=0.05$, $\sigma_1=\sigma_2=0.3$, $\tau=1$.}
\label{fig: price-floating}
\end{figure}

\begin{figure}[!htbp]
\begin{center}
\includegraphics[width=0.49\linewidth]{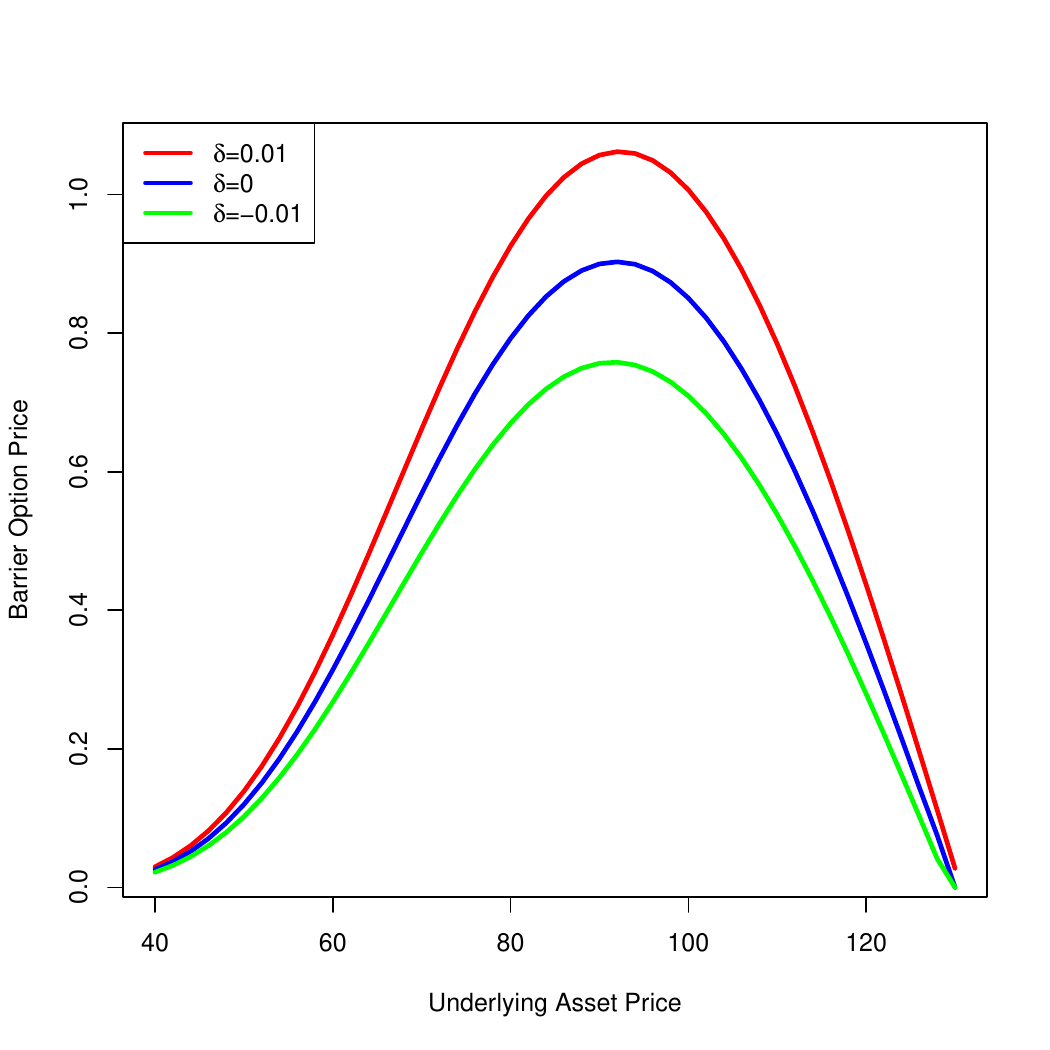}
\includegraphics[width=0.49\linewidth]{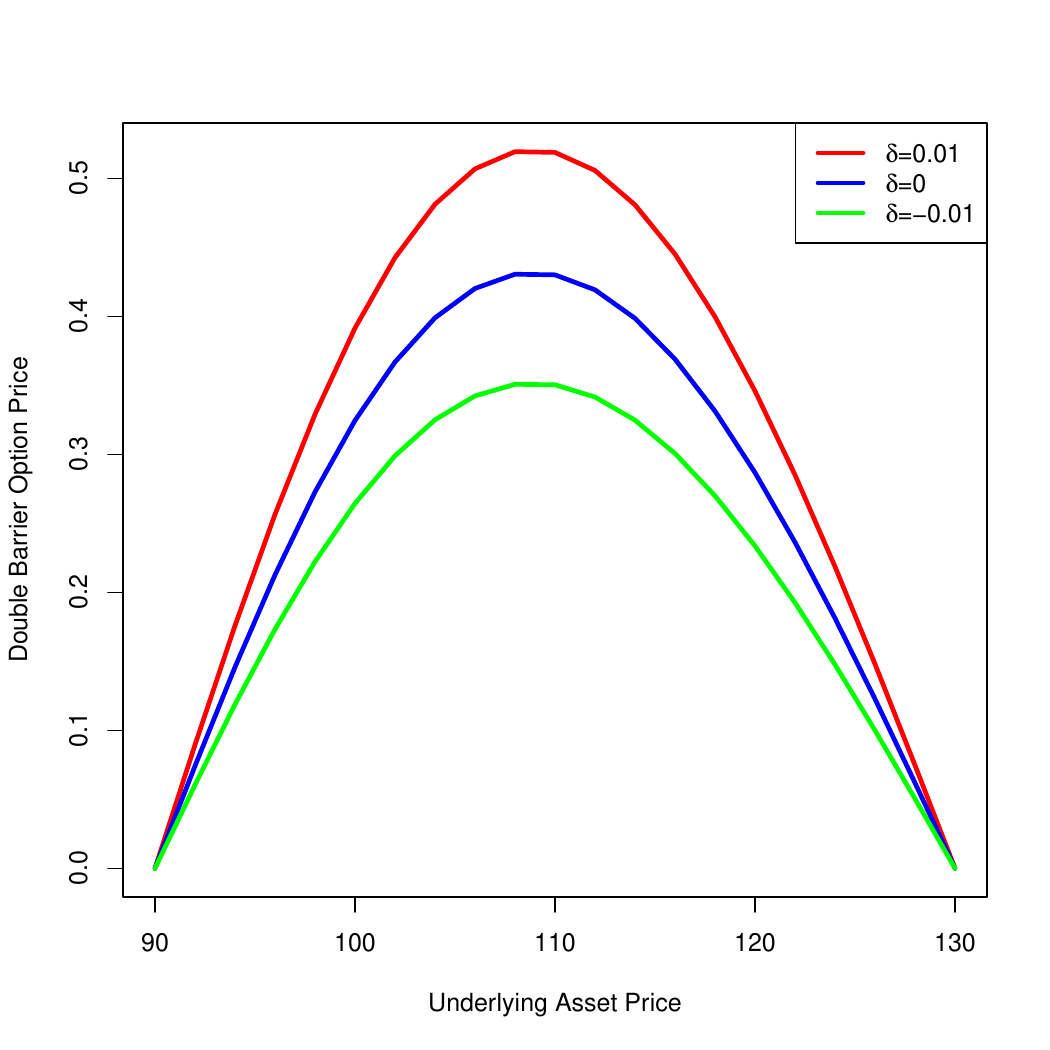}
\end{center}
\caption{Single barrier up-and-out call price (left) and  double barrier up-and-out call price (right) as functions of underlying price for different floating rates $\delta$. Parameters: $S_{B_1}=100$, $S_B=S_{B_2}=130$, $K=100$, $\rho=0.5$, $a=1$, $\theta=0.04$, $r_0=0.05$, $\sigma_1=\sigma_2=0.3$, $\tau=1$.}
\label{fig: price-underlying}
\end{figure}

\begin{figure}[!htbp]
\begin{center}
\includegraphics[width=0.49\linewidth]{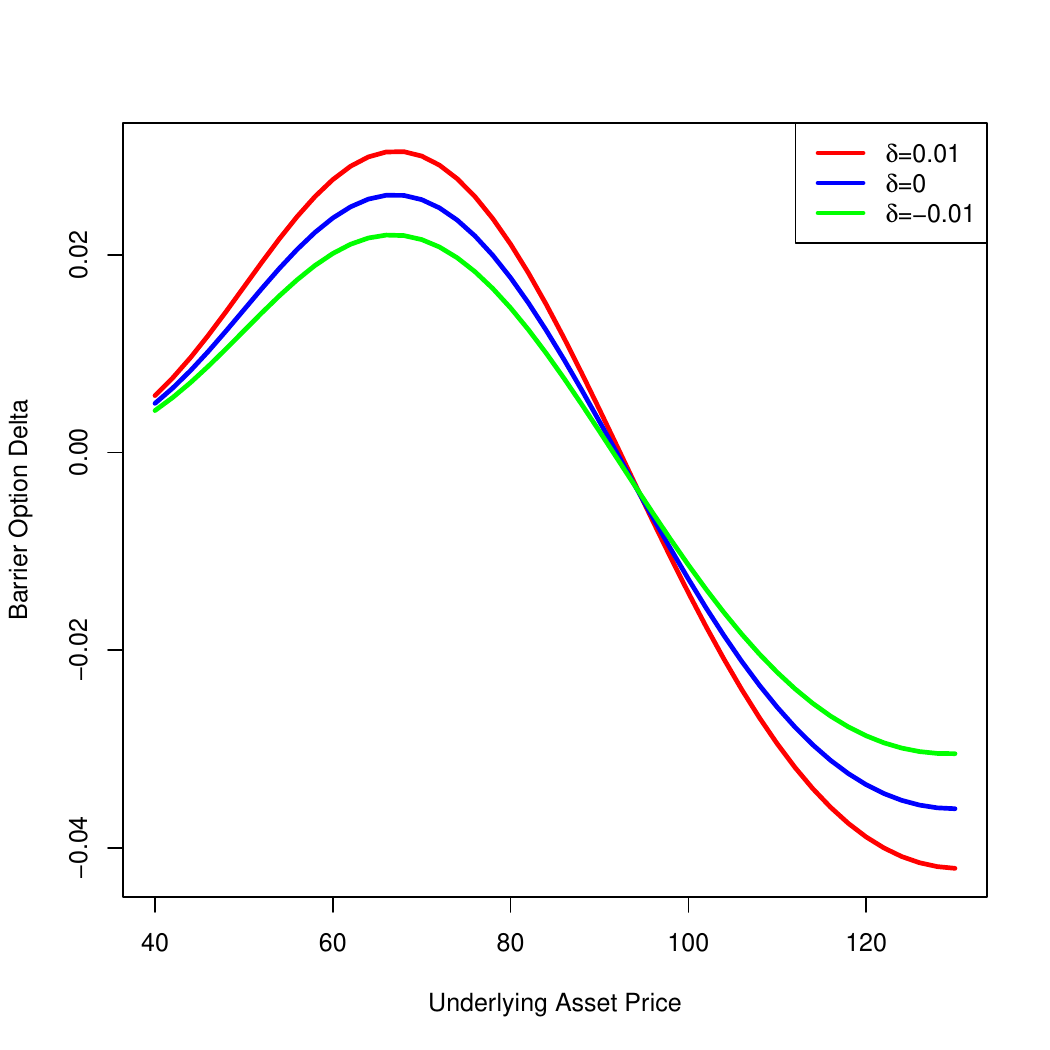}
\includegraphics[width=0.49\linewidth]{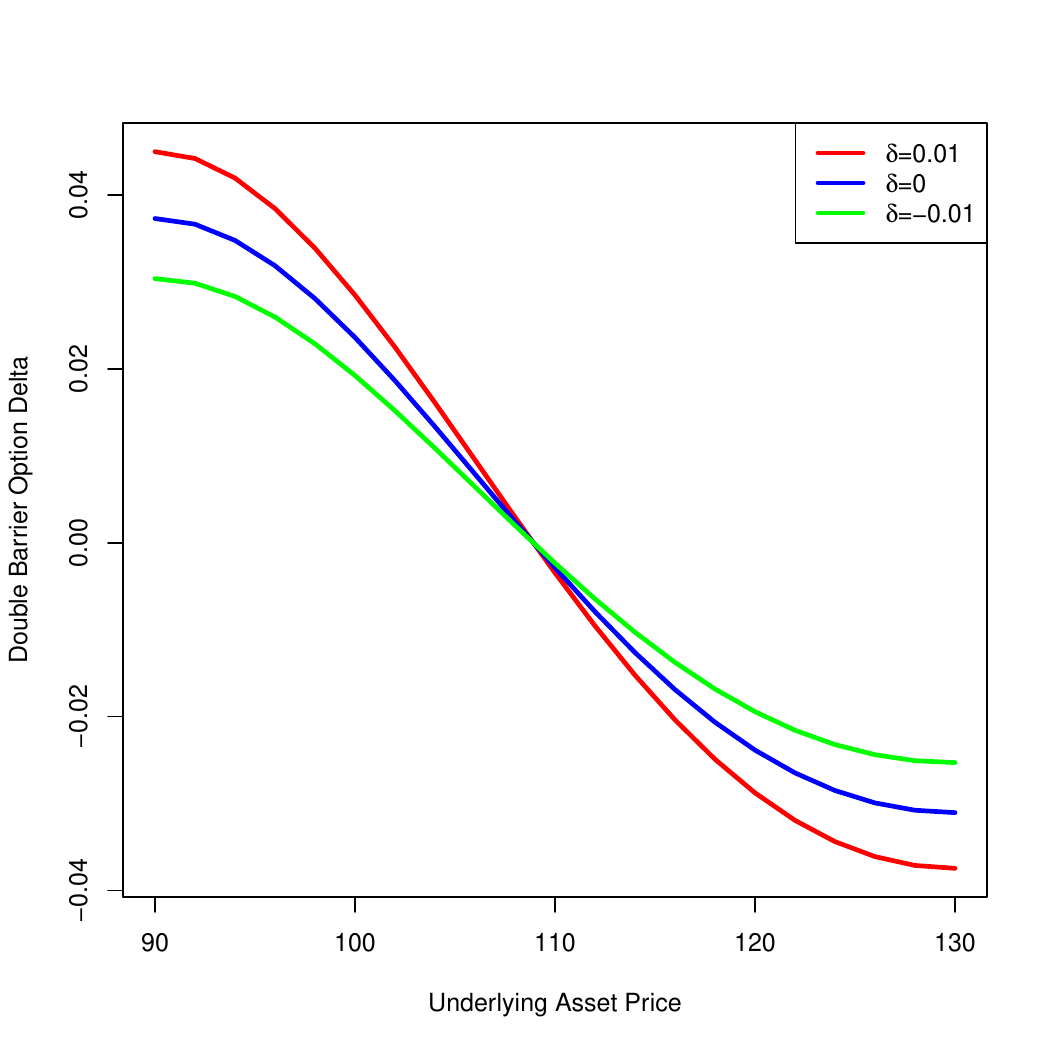}
\end{center}
\caption{Barrier call Delta (left) and  double-barrier call Delta (right) as functions of underlying price for different floating rates. Parameters: $S_{B_1}=100$, $S_B=S_{B_2}=130$, $K=100$, $\rho=0.5$, $a=1$,$\theta=0.04$, $r_0=0.05$, $\sigma_1=\sigma_2=0.3$, $\tau=1$.}
\label{fig: Delta-underlying}
\end{figure}

%%%%%%%%%%%%%%%%%%%%%%%%%%%%%%%%%%%%%%%%%%%%%%%%%%%%%%%%%%%%%%%%%%%%%%%%%%%%%%%%%%%%
Table.~\ref{TABLE1} shows the dependence of a up-and-out barrier call price and a up-and-out double barrier call price on the  underlying price for different barrier floating rates $\delta$. $r_0=0.05$ is the initial interest rate. Without loss of generality, we set $\sigma_1=\sigma_2=0.3$. The underlying asset price $S$ and the barrier level $B$($B_2$) are joint to determine the option price. When $S$ is far smaller than the exercises price $K$, $S$ dominates the option price and there tends to be less probability to exercise the option. Hence the option price is very low. When $S$ is near the barrier, the option has more probabilities to touch the barrier, and the option price is low as well. In Fig.~\ref{fig: price-floating}, we show the dependence of the option price on the floating rate $\delta$ for different underlying prices. With the increase of $\delta$, the option price increases. In Fig.~\ref{fig: price-underlying}, it is obviously shown that the price of the option with a positive $\delta$ (the red line) is higher than that of a standard barrier option with the same exercise price $K$, since the option has more space to survive. For the same reason, the option with a negative $\delta$ (the green line) is cheaper than the standard barrier option with the same $K$. The dependence of option Delta on the underlying asset price is shown in Fig.~\ref{fig: Delta-underlying}.

\begin{table}[h]
\centering
\begin{tabular}{|c|c|c|c|c|c|c|}
\hline
\multirow{2}*{Underlying Price $S$} & \multicolumn{3}{c|}{Single Barrier Option Price $C_{\rm B}$} & \multicolumn{3}{c|}{Double Barrier Option Price $C_{\rm DB}$}\\
\cline{2-7}
    & $\delta=-0.01$ & $\delta=0$ & $\delta=0.01$ & $\delta=-0.01$ & $\delta=0$ & $\delta=0.01$ \\
\hline
  40   & 0.0218088 & 0.0255593 & 0.0226368 & - & - & -  \\
  60   & 0.267311 & 0.31392 & 0.364335 & - & - & -  \\
  80   & 0.669909 & 0.792455 & 0.925853 & - & - & -  \\
  90   & 0.756554 & 0.89997 & 1.0571 & 0 & 0 & 0  \\  
  100   & 0.70949 & 0.850758 & 1.00701 & 0.26442 & 0.324539 & 0.391472  \\
  120   & 0.275938 & 0.352539 & 0.437257 & 0.233614 & 0.286748 & 0.345922 \\
  128  & 0.041192 & 0.0748869& 0.111453 &0.0504391 & 0.061912 & 0.0746894 \\
  130  & 0 & 0 & 0 & 0 & 0 & 0 \\
 \hline
\end{tabular}
\caption{Single barrier up-and-out call price $C_{\rm B}$ and double barrier up-and-out call option price $C_{\rm DB}$ as functions of the underlying asset price $S$ under different floating rates $\delta$. Parameters: $S_{B_1}=100$, $S_{B}=S_{B_2}=130$, $K=100$, $\rho=0.5$, $a=1$, $\theta=0.04$, $r_0=0.05$, $\sigma_1=\sigma_2=0.3$, $\tau=1$. }
\label{TABLE1}
\end{table}

\begin{figure}[!htbp]
\begin{center}
\includegraphics[width=0.49\linewidth]{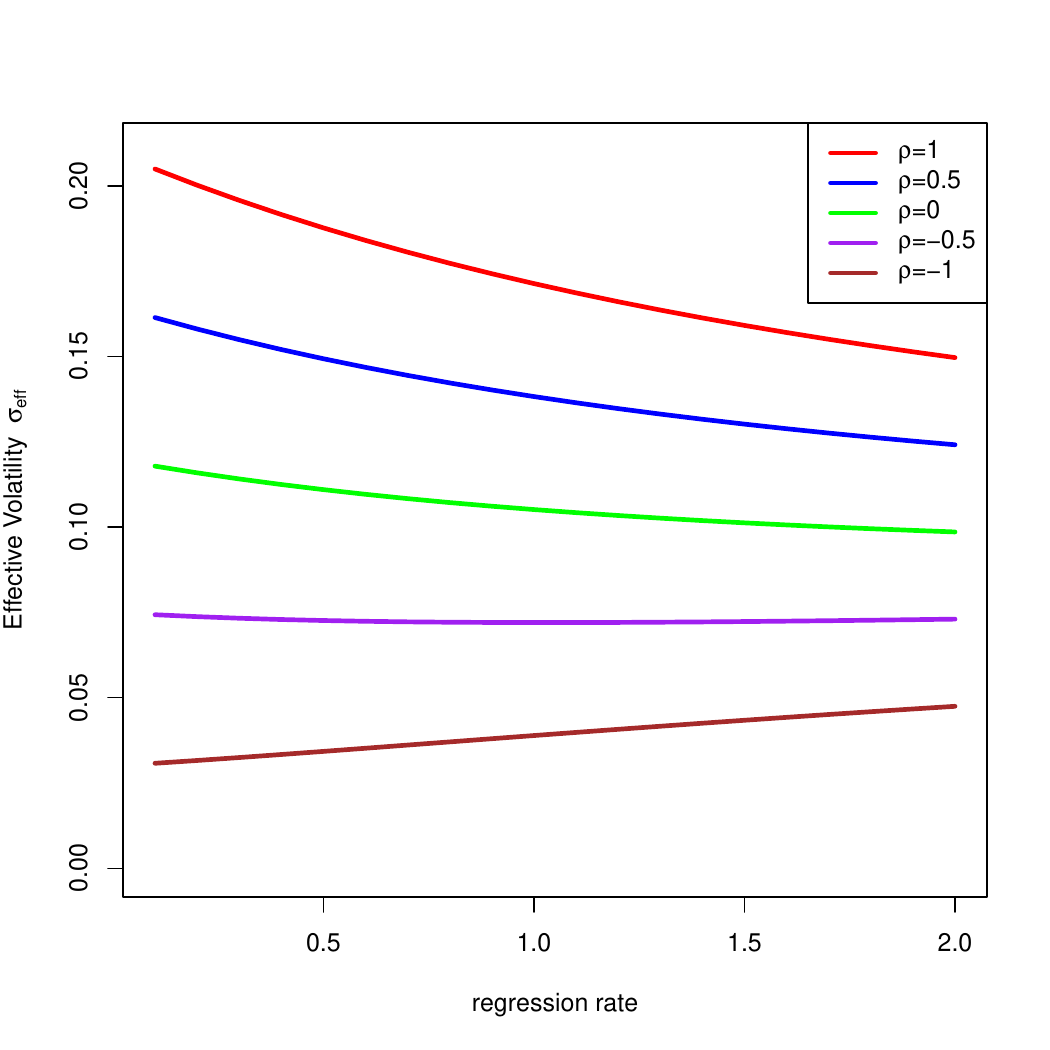}
\end{center}
\caption{Effective volatility $\sigma_{\rm eff}$ as a function of regression rate $a$ for different $\rho$s. Parameters: $\sigma_1=\sigma_2=0.3$, $\tau=1$.}
\label{fig: price-effsigma}
\end{figure}

\begin{figure}[!htbp]
\begin{center}
\includegraphics[width=0.49\linewidth]{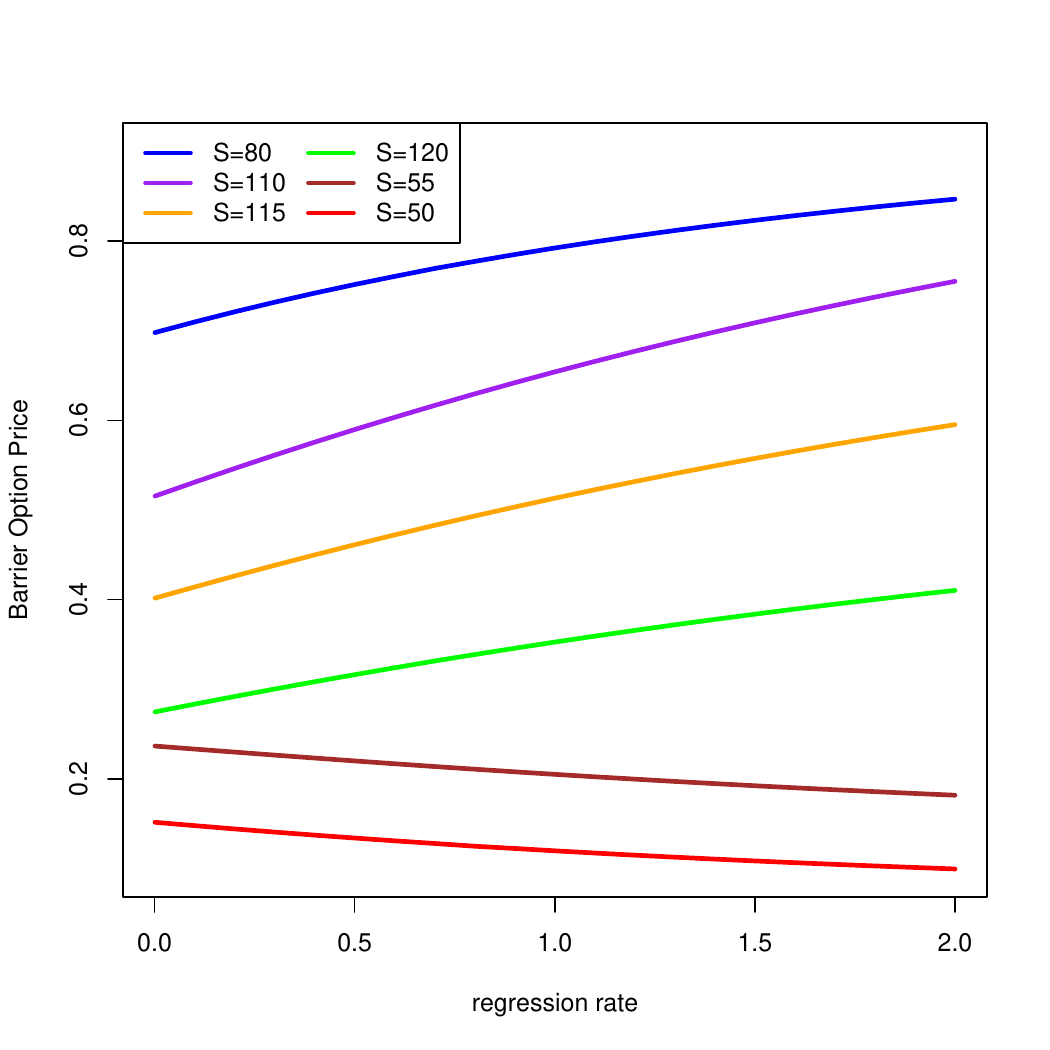}
\includegraphics[width=0.49\linewidth]{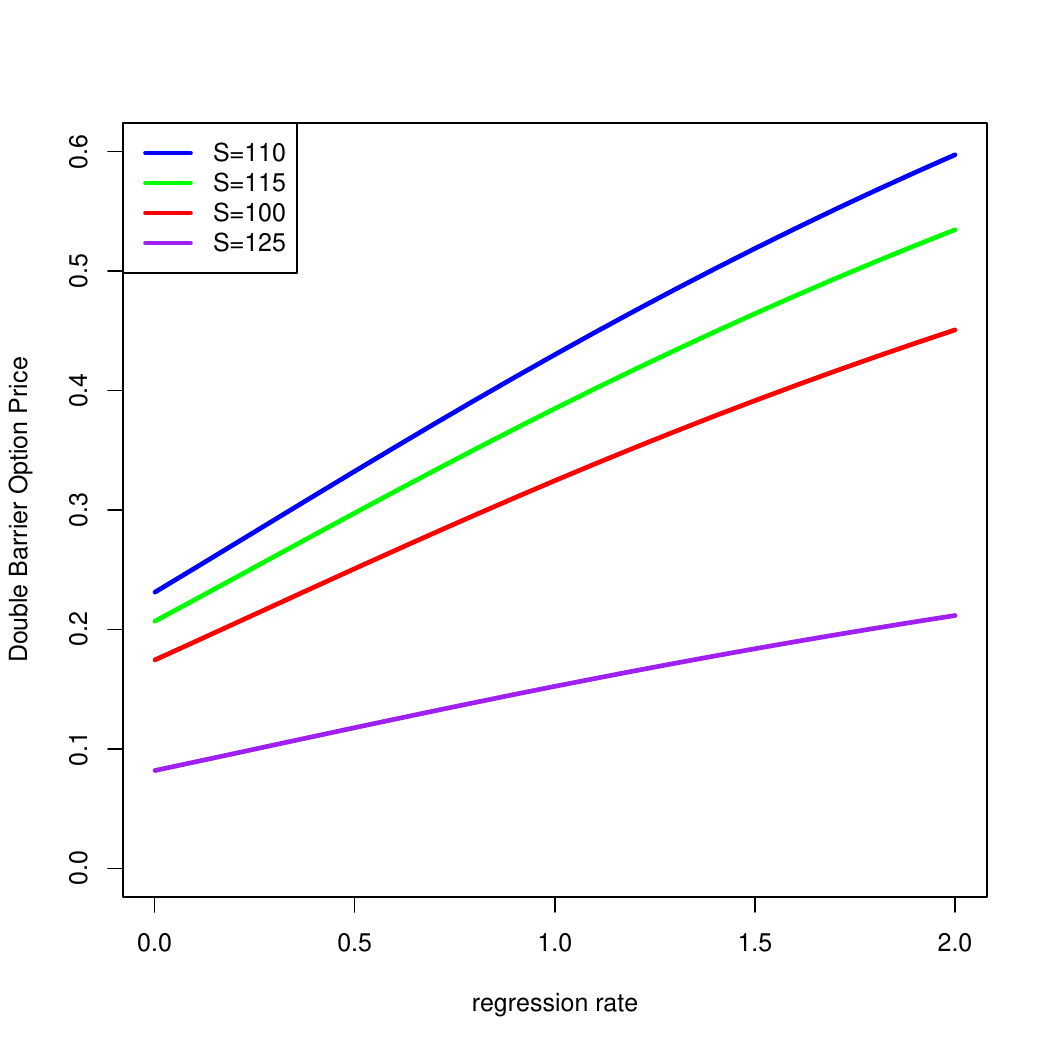}
\end{center}
\caption{Barrier call price (left) and  double-barrier call price (right) as functions of regression rate $a$ for different underlying asset prices. Parameters: $S_{B_1}=100$, $S_B=S_{B_2}=130$, $K=100$, $\rho=0.5$, $\theta=0.04$, $\delta=0$, $r_0=0.05$, $\sigma_1=\sigma_2=0.3$, $\tau=1$.}
\label{fig: price-regression}
\end{figure}

In Fig.~\ref{fig: price-regression}, we show the dependence of the option price on the regression rate $a$ for different underlying asset prices at $\rho=0.5$. For a single barrier option, it is shown that in Fig.~\ref{fig: price-effsigma}, the effective volatility $\sigma_{\rm eff}$ decreases with the increase of $a$ for $\rho=0.5$, hence the option price decreases with the increasing $a$ at a low underlying price. When $S$ is near the barrier level, the decrease of $\sigma_{\rm eff}$ would make the option has less opportunities to touch the barrier, and the option price increases with the increasing $a$. For a double barrier option, the decrease of $\sigma_{\rm eff}$ makes the option has less opportunities to touch both sides of the barriers, and the option price increases with the increase of $a$.

%%%%%%%%%%%%%%%%%%%%%%%%%%%%%%%%%%%%%%%%%%%%%%%%%%%%%%%%%%%%%%%%%%%%%%%%%%%%%%%%%%%%%%
\section{Conclusion}
%%%%%%%%%%%%%%%%%%%%%%%%%%%%%%%%%%%%%%%%%%%%%%%%%%%%%%%%%%%%%%%%%%%%%%%%%%%%%%%%%%%%%

Barrier option price changing with time could be analogous to a particle moving under some special potential. For stochastic interest rates, the maturity should be split into $N$ steps, and each matrix element during a tiny step is calculated by Gaussian integral. After $N$ times integral calculation, the pricing kernel could be written in the form of a series, which could be represented as some special function. Hamiltonian is an effective approach linking option price changing to a particle moving under some potential in the space. The pricing of other barrier options such as step options could be studied by defining appropriate potentials. 
\section*{Acknowledgments}
 Chao Guo and Ning Yao are supported by the Natural Science Foundation of Higher Education Institutions in Hebei Province under Grant No. QN2023259.

%%%%%%%%%%%%%%%%%%%%%%%%%%%%%%%%%%%%%%%%%%%%%%%%%%%%%%%%%%%
\appendix
%%%%%%%%%%%%%%%%%%%%%%%%%%%%%%%%%%%%%%%%%%%%%%%%%%%%%%%%
\section{Hamiltonian Approach to the Standard Barrier Option Pricing}
%%%%%%%%%%%%%%%%%%%%%%%%%%%
The up-and-out standard barrier (UOSB) option Hamiltonian is
\begin{equation}\begin{split}
H_{\rm UOSB}&=H_{\rm BS}+V(x)\\
&=-\frac{\sigma^2}{2}\frac{\partial^2}{\partial x^2}+\left(\frac{\sigma^2}{2}-r\right)\frac{\partial}{\partial x}+r+V(x)\\
&=e^{\alpha x}\left(-\frac{\sigma^2}{2}\frac{\partial^2}{\partial x^2}+\gamma\right)e^{-\alpha x}+V(x)
\end{split}\end{equation}
where 
\begin{equation}
    \alpha=\frac{1}{\sigma^2}\left(\frac{\sigma^2}{2}-r\right),\ \ \gamma=\frac{1}{2\sigma^2}\left(\frac{\sigma^2}{2}+r\right)^2
\end{equation}
and the potential $V(x)$ is
\begin{equation}
V(x)=\left\{
\begin{aligned}
0 & , & x< B,\\
\infty & , & x\geq B.
\end{aligned}
\right.
\end{equation}
the corresponding wave function is
\begin{equation}
 C(x)=\left\{
\begin{aligned}
&e^{ip(x-B)}-e^{-ip(x-B)}  , & x<B,\\
&0  , & x\geq B.
\end{aligned}
\right.   
\end{equation}
and the pricing kernel is
\begin{equation}\begin{split}
p_{\rm UOSB}(x,x^\prime;\tau)&=\braket{x|e^{-\tau H_{DB}}|x^\prime}\\
&=e^{-\tau\gamma}e^{\alpha(x-x^\prime)}\int_{0}^{\infty}\frac{{\rm d}p}{2\pi}e^{-\frac{1}{2}\tau\sigma^2p^2}\big[e^{ip(x-B)}-e^{-ip(x-B)}\big]\big[e^{-ip(x^\prime-B)-e^{ip(x^\prime-B)}}\big]\\
&=2e^{-\tau\gamma}e^{\alpha(x-x^\prime)}\int_{0}^{\infty}\frac{{\rm d}p}{2\pi}e^{-\frac{1}{2}\tau\sigma^2p^2}[\cos(p(x-x^\prime))-\cos(p(x+x^\prime-2B))]
\end{split}\end{equation}
the corresponding option price
\begin{equation}\begin{split}
    C_{\rm UOSB}(x;\tau)&=\int_{{\rm ln}K}^B {\rm d}x^\prime p_{\rm UOSB}(x,x^\prime;\tau)(e^{x^\prime}-K)\\
    &=2e^{-\tau\gamma}\int_{{\rm ln}K}^B {{\rm d}x^\prime}e^{\alpha(x-x^\prime)}\int_0^{\infty}\frac{{\rm d}p}{2\pi}e^{-\frac{1}{2}\tau\sigma^2p^2}[\cos(p(x-x^\prime))-\cos(p(x+x^\prime-2B))](e^{x^\prime}-K)
\end{split}\end{equation}

%%%%%%%%%%%%%%%%%%%%%%%%%%%
\section{Hamiltonian Approach to the Standard Double-Barrier Option Pricing}
%%%%%%%%%%%%%%%%%%%%%%%%%%%
The standard double-barrier (SDB) option Hamiltonian is~\cite{Baaquie}
\begin{equation}\begin{split}
    H_{\rm SDB}&=H_{\rm BS}+V(x)\\
    &=-\frac{\sigma^2}{2}\frac{\partial^2}{\partial x^2}+\left(\frac{\sigma^2}{2}-r\right)\frac{\partial}{\partial x}+r+V(x)\\
&=e^{\alpha x}\left(-\frac{\sigma^2}{2}\frac{\partial^2}{\partial x^2}+\gamma\right)e^{-\alpha x}+V(x)
\end{split}\end{equation}
where 
\begin{equation}
    \alpha=\frac{1}{\sigma^2}\left(\frac{\sigma^2}{2}-r\right),\ \ \gamma=\frac{1}{2\sigma^2}\left(\frac{\sigma^2}{2}+r\right)^2
\end{equation}
and the potential $V(x)$ is
\begin{equation}
V(x)=\left\{
\begin{aligned}
\infty & , & x\leq a,\\
0 & , & a<x<b,\\
\infty & , & x\geq b.
\end{aligned}
\right.
\end{equation}
the corresponding eigenstate is
\begin{equation}
 \phi_n(x)=\left\{
\begin{aligned}
\sqrt{\frac{n\pi}{b-a}}\sin&[p_n(x-a)]  , & a<x<b,\\
0 & , & x<a,\ x>b.
\end{aligned}
\right.   
\end{equation}
where 
\begin{equation}
    p_n=\frac{n\pi}{b-a},\ \ E_n=\frac{1}{2}\sigma^2p_n^2,\ \ n=1,2,3,...
\end{equation}

The pricing kernel is
\begin{equation}\begin{split}
    p_{\rm SDB}(x,x^\prime;\tau)&=\braket{x|e^{-\tau H_{DB}}|x^\prime}\\
    &=e^{\alpha(x-x^\prime)}\braket{x|e^{-\tau\left(-\frac{\sigma^2}{2}\frac{\partial^2}{\partial x^2}+\gamma+V\right)}|x^\prime}\\
    &=e^{-\tau\gamma}e^{\alpha(x-x^\prime)}\sum_{n=1}^{+\infty}e^{-\frac{1}{2}\tau\sigma^2p_n^2}\phi_n(x)\phi_n(x^\prime)
\end{split}\end{equation}
and the option price 
\begin{equation}\begin{split}
C_{\rm SDB}(x;\tau)&=\int_{{\rm ln}K}^b{\rm d}x^\prime p_{\rm SDB}(x,x^\prime;\tau)(e^{x^\prime}-K)\\
&=\frac{2}{b-a}e^{-\tau\gamma}\int_{{\rm ln}K}^b{\rm d}x^\prime e^{\alpha(x-x^\prime)}\sum_{n=1}^{+\infty}e^{-\frac{1}{2}\tau\sigma^2\frac{n^2\pi^2}{(b-a)^2}}\sin\frac{n\pi(x-a)}{b-a}\sin\frac{n\pi(x^\prime-a)}{b-a}(e^{x^{\prime}}-K)
\end{split}\end{equation}

%%%%%%%%%%%%%%%%%%

\end{document}